\documentclass[aps,twocolumn,secnumarabic,nobalancelastpage,amsmath,amssymb,nofootinbib]{revtex4}

\usepackage{graphics}      % standard graphics specifications
\usepackage{graphicx}      % alternative graphics specifications
\usepackage{longtable}     % helps with long table options
\usepackage{url}           % for on-line citations
\usepackage{bm}            % special 'bold-math' package
\begin{document}
\title{A study for the energy structure of the Mott system with a low-energy excitation, in terms of the pseudogap in HTSC.\\}
\author         {Keishichiro Tanaka$^1$}
\email          {keishichiro.tanaka@gmail.com\
ORCiD:https://orcid.org/0000-0002-9995-0959}
%\homepage       { }
\affiliation    {}
\date{\today}

%abstract%%%%%%%%%%%%%%%%%%%%%%%%%%%%%%%%%%%%%%%%%%%%%%%%%%%%%%%%%%%%%%%%%%%%%%%%%%
\begin{abstract}
This paper shows that the origin of the pseudogap in high-temperature superconductivity (HTSC) cuprates is attributed to the self-energy effect on a quasiparticle excitation of approximately \begin{math} 4t^2/U \end{math} per two electrons, which is the Mott's $J$. The pseudogap of HTSC is a phenomenon that has remained unexplained since the discovery of HTSC. This paper examines the Mott system with a low-energy excitation as the underlying system for HTSC in under-doped cuprates, in order to elucidate the phenomenon, through the Hubbard-1 approximation (Green's function method), especially in terms of the self-energy effect.
\end{abstract}
\maketitle

%%%%%%
%Introduction%%%%%%%%%%%%%%%%%%%%%%%%%%%%%%%%%%%%%%%%%%%%%%%%%%%%%%%%%%%%%%%%%%%%%%%%
%%%%%%
\section{Introduction}

This paper seeks to identify the origin of the pseudogap in high-temperature superconductivity (HTSC) cuprates, which has remained unexplained since the discovery of HTSC. In this respect, this paper may contribute to elucidating the mechanisms of HTSC cuprates and other unconventional superconductors.

The mechanism of HTSC is considered to differ from that described by the BCS theory \cite{reference1, reference2, reference3}. The superconducting state of HTSC cuprates emerges near Mott insulators, and features two types of energy gaps --- a superconductivity-gap and a pseudogap appearing as a suppression of electronic spectral weight --- \cite{reference4, reference5, reference6, reference7, reference8} as well as the Fermi surface reconstruction \cite{reference9, reference10, reference11}, as observed in spectroscopic experiments such as STS and ARPES. So far, various theories for HTSC have been proposed, including those based on spin fluctuations \cite{reference12, reference13}, density waves \cite{reference14, reference15}, spin-charge separation \cite{reference16}, two-component fermions \cite{reference17}, multi-orbital effects \cite{reference18, reference19}, and super-current by spin twisting \cite{reference20}. In addition, a variety of other types of superconductors have been discovered \cite{reference21,reference22,reference23}, in which low-energy excitations have often been key elements in the discussions. 

%%%%%%%%%%%%%%%%%%%%%%%%%%%%%%%%%%%%%%%%%%%%%%%%%%%%%%%%%%%%%%%%%%%%%%%%% 04/11/25
Here, we review some previous studies on the pseudogap (PG) mechanisms that are particularly thought to be related to ordered phases in under-doped cuprates, spin density waves (SDWs) and charge density waves (CDWs), both of which are based on stable theoretical frameworks. In principle, SDW and CDW compete with superconductivity (SC) in terms of U(1) gauge symmetry. Recently, other experimental results suggest that the PG is an electronic nematic phase, which has translational symmetry but breaks rotational symmetry \cite{reference24}. However, it should be noted that the competitive relationship between CDW/SDW and SC in this system is still under debate \cite{reference25}. 

As for SDWs, several studies have been conducted to incorporate thermal fluctuations into the correction of the self-energy and/or the order parameter of the SDW in the mean-field approximation \cite{reference26,reference27}. These studies indicate the similarity between the calculated spectrum of the SDW and the ARPES spectrum of the PG, as well as the Fermi surface restoration due to thermal fluctuations at higher temperature, and conclude that the PG is a precursor of the SDW, based on the analyzation of the spectral functions. However, according to experimental results, the SC and the PG appear at higher temperatures and over a broader doping range than the SDW appears in the recent phase diagram \cite{reference25}, and SDWs have not yet been clearly observed in the $\textbf{\textit{q}}$ = (0,1)/(1,0) direction in under-doped cuprates. 

As for CDWs, there is a study that successfully calculates the direction of the CDW (along the $\textbf{\textit{q}}$ = (0,1)/(1,0)), based on the instability of the CDW susceptibility due to a smaller short-range coulomb repulsion $U(\textbf{\textit{q}})$ by using the Lindhardt polarization function, as well as its onset temperature by using the mass gap related to quantum critical point (QCP) in YBCO. The latter shows the calculated onset temperature of the CDW and the experimental onset temperature of the PG are in good agreement, and then concludes the PG is a phenomenon related to a CDW mediated fluctuation \cite{reference28}. It is understandable that the onset of the SC would be enhanced as the CDW-meditated fluctuations become stronger. However, the process by which the fluctuations due to the CDW form an additional energy gap (PG) is not shown in their model, though this may be related to what they call an internal d-wave structure not incorporated in their CDW order parameter. In addition according to experimental results, the CDW and its CDW-mediated fluctuations that they hypothesize for the formation of the PG should disappear when the SC appears \cite{reference29}, but in fact the PG is still there in under-doped cuprates. Alternatively, the above coincidence may represent different phenomena arising from the same cause, $U(\textbf{\textit{q}})$. 

As a whole, most of the studies on the pseudogaps of HTSC to date have attempted to explain the entire mechanism of high-temperature superconductivity somewhat qualitatively, by assuming the connection between the pseudogap and the ordered phase and by treating the pseudogap as a precursor phenomenon of the ordered phase or Cooper pairs, mainly using the mean-field approximation. These have revealed many interesting phenomena of HTSC. However, the consistency of the SC, PG and competing ordered phases with the phase diagram and how the fluctuations themselves open up energy gaps have not been clearly shown \cite{reference25}. By contrast, the present study mainly focuses on a quantitative evaluation of the pseudogap itself in a strong-correlated system and try to explore its intrinsic magnitude, by calculating an additional self-energy more directly at the anti-nodal point, which is another high symmetry point, as well as the local self-energy (on-site), while the dispersion relation is unclear. Also this study basically develops the discussion at \begin{math} T=0 \end{math}, where the thermal fluctuations of the parameters are stable, and only confirms the tendency of the pseudogap behavior at finite temperatures. This study is attempting to to adopt a different approach to the elucidation of the pseudogaps in HTSC.
%%%%%%%%%%%%%%%%%%%%%%%%%%%%%%%%%%%%%%%%%%%%%%%%%%%%%%%%%%%%%%%%%%%%%%%%% 04/11/25

First, this study explains a low-energy excitation, $\Delta_{J}$ (and $\delta$) to be discussed, using the effective Hamiltonian. Second, it evaluates the self-energy effects on this excitation by means of the Hubbard-1 approximation, the Green's function method that represents the Mott system incorporating self-energy \cite{reference3}\cite{reference30,reference31,reference32,reference33,reference34}. Third, while verifying the estimated self-energy with the effective-mass, it compares the resulting values of the Green's function poles with experimental pseudogap values in Bi2212s.

Typically, the Hubbard-1 approximation first determines the self-energy in the atomic limit, and then incorporates dispersion relations \begin{math} \epsilon (\textbf{\textit{k}}) \end{math} when extending to a band structure \cite{reference33, reference34}. In contrast, this study incorporates a specific excitation of the Mott system into the approximation, instead of $\epsilon (\textbf{\textit{k}})$. This excitation is assumed to appear in the direction of (0, $\pi$) and ($\pi$, 0) in $k$-space (reciprocal lattice space) in under-doped HTSCs \cite{reference7} and is considered to be a quasiparticle excitation associated with the antiferromagnetic interaction, $J$(\begin{math} \sim 4t^2/U \end{math}), of the Mott system.
%%%%%%%%%%%%%%%%%%%%%%%%%%%%%%%%%%%%%%%%%%%%%%%%%%%%%%%%%%%%%%%%%%%%%%%%%%%%%%%%%

%%%%%%
%Effective Hubbard model%%%%%%%%%%%%%%%%%%%%%%%%%%%%%%%%%%%%%%%%%%%%%%%%%%%%%%%%%%%%%%%%%%%%
%%%%%%
\section{Effective Hamiltonian}

This section explains the meaning of a low-energy excitation of the Mott system, $\Delta_{J}$ (and $\delta$), which is the subject of the discussion in this study, using the effective Hamiltonian as outlined below. 
%%%%%%
\footnotesize
\begin{subequations}
\begin{eqnarray} \label{eq:eq1} \label{eq:Hubbard_hamiltonian_normal_state}
H_{eff}^{}&=&-t\sum_{<i,j>\sigma} \left( c^{\dagger}_{r_{i} \sigma} c^{}_{r_{j} \sigma^{}}+ c^{\dagger}_{r_{j} \sigma} c^{}_{r_{i} \sigma^{}}\right) \nonumber\\
&+&\Delta_{J} \sum_{(i)}c^{\dagger}_{r_{i} \sigma} c^{}_{r_{i} \sigma^{}} \nonumber\\
&+& U \sum_{(i)}^{N} n_{i\uparrow}n_{i\downarrow}.
\end{eqnarray}
\end{subequations}
\normalsize
%%%%%%
where $t$ is the transfer integral, $U$ is the coulomb repulsion, and $\Delta_{J}$ is an original low-energy excitation per electron specific to this Mott system, which appears in the direction of (hereafter, 'at' will be used) \begin{math} \textbf{\textit{k}}  = (k_{x},k_{y}) = (0,\pi)/(\pi, 0) \end{math} in k-space, which is a high symmetry point in terms of the lattice symmetry. $r_{i} \sigma$ and $r_{j} \sigma$ represent a pair of nearest-neighbor sites with spin $\sigma$ (=$\uparrow$ or $\downarrow$), $n_{i} \sigma$ is the density operator. The second term with $\Delta_{J}$ simply represents a non-interacting excitation term in this paper, but it may actually be an interaction term. In this study, it is assumed that \begin{math} t=0.4eV \end{math}, \begin{math} U=3.2eV \end{math}, and \begin{math} U/t =8 \end{math} for an occupation ratio \begin{math} \nu=0.8 \end{math}. The self-energy effect of this system shifts $\Delta_{J}$ to $\delta$ which is assumed to be a pseudogap of HTSC per electron, demonstrated in Section 3.\\
%%%%%%%%%%%%%%%%%%%%%%%%%%%%%%%%%%%%%%%%%%%%%%%%%%%%%%%%%%%%%%%%%%%%%%%%%%%%%%%%%

%%%%%%
%Green function for one particle%%%%%%%%%%%%%%%%%%%%%%%%%%%%%%%%%%%%%%%%%%%%%%%%%%%%%%%%%%%%%%%%%
%%%%%%
\section{Green's functions}

\subsection{Hubbard-1 approximation}
This study evaluates the excitation energy $\Delta_{J}$ and its shifted energy $\delta$ per electron in the effective Hamiltonian in Eq.(1a) at \begin{math} \textbf{\textit{k}} = (0,\pi)/ (\pi, 0) \end{math} using the Hubbard-1 approximation \cite{reference3} \cite{reference33, reference34}, non-interacting single-particle Green's function method incorporating the self-energy in the atomic limit (\begin{math} t = 0 \end{math}).
% This approximation analyzes the poles of its Green's functions using the self-energy in the atomic limit (\begin{math} t = 0 \end{math}).
% The Green's function and the self-energy here are divided into two parts, $G^{R}_{1} (\textbf{}^{}\omega)_{}$ and $\Sigma^{R}_{1} (\textbf{}^{}\omega)$ (local self-energy) per electron, which are independent of \textbf{k} \cite{reference27} usually at \begin{math} \textbf{k}_{F} \end{math}, as well as $G^{R}_{2} (\textbf{}^{}\omega)_{}$ and $\Sigma^{R}_{2} (\textbf{}^{} \omega^{})$ per electron, which relate to a quasi-particle excitation $\Delta_{J}$ at \begin{math} \textbf{\textit{k}} = (0,\pi)/ (\pi, 0) \end{math}. 

In this study, the Green's function for the approximation incorporates the self-energy at \begin{math} \textbf{\textit{k}} = (0,\pi)/ (\pi, 0) \end{math}, which is another high-symmetry point, where an excitation $\Delta_{J}$ is observed, in addition to the local self-energy (on-site) of the system, which is independent of \textbf{\textit{k}} \cite{reference33}. The local energy here refers to the energy that acts over a small range.
%The local energy here refers to energy that may not inherently reflect the periodicity of the system.

To begin with, this approximation calculates the local self-energy (on-site) $\Sigma^{R}_{1} (\textbf{}^{}\omega)$ per electron using Eqs.(2). Next, it calculates the self-energy $\Sigma^{R}_{2} (\textbf{}^{}\omega)$ per electron at \begin{math} \textbf{\textit{k}}= (0,\pi)/ (\pi, 0) \end{math} using Eqs.(3). Finally, it constructs the Green's function incorporating both $\Sigma^{R}_{1} (\textbf{}^{}\omega)$ and $\Sigma^{R}_{2} (\textbf{}^{}\omega)$ per electron as shown in Eqs.(6), in order to estimate the magnitudes of the energy excitations, $\Delta_{J}$ and $\delta$, per electron at \begin{math} \textbf{\textit{k}} = (0,\pi)/ (\pi, 0) \end{math}. 

Although the Hubbard-1 approximation does not account for the \textbf{\textit{k}}-dependent self-energy, this study assumes that the same logic for the local self-energy (on-site) in Eqs.(2) can apply to conditions near the atomic limit, such as that at \begin{math} \textbf{\textit{k}} = (0,\pi)/ (\pi, 0) \end{math} of HTSC, where an excited structure is clearly observed. Then if applicable, the self-energy at \begin{math} \textbf{\textit{k}} = (0,\pi)/ (\pi, 0) \end{math} of the system represents a type of local-energy.

The following Green's functions build on the thermal average of the particle number based on the Fermi-Dirac distribution. 
%As a quasi-particle excitation with a finite life time, what an electron hops into the $\Delta_{J}$ level is possible even at lower temperatures than that of its thermal excitation on average. 
%%%%%%

%%%%%%
% Single-particle Green's function %%%%%%%%%%%%%%%%%%%%%%%%%%%%%%%%%%%%%%%%%%%%%%%%%%%%%%%%%%%%%%%%%%%%
%%%%%%
\subsection{Green's function and local self-energy }
The single-particle Green's function ($G^{R}_{1} (\omega)_{}$) and its local self-energy (on-site) ($\Sigma^{R}_{1} (\textbf{}^{}\omega)$) in the atomic limit, which is independent of \textbf{\textit{k}}, derived through the Dyson equation are shown as follows. In this system, however, $G^{R}_{1} (\omega)_{}$ corresponds to the Green's function at \begin{math} \textbf{\textit{k}}^{} = \textbf{\textit{k}}_{F} \end{math}, the Fermi wave vector, from an energy perspective.
%%%%%%
\footnotesize
\begin{subequations}
\begin{eqnarray} \label{eq:eq2}
%G^{R}_{} (\omega)_{}&=&G^{R}_{1} (\textbf{}^{}\omega)_{}+G^{R}_{2} (\textbf{}^{}\omega)_{}. \\
G^{R}_{1} (\textbf{}^{}\omega)_{}&=&\frac{1-<n_{-\sigma, \bar{\Delta_{J}}}>} {\omega+\mu^{'}-0} + \frac{<n_{-\sigma, \bar{\Delta_{J}}}>} {\omega+\mu^{'}-U}. \\
%\Sigma^{R}_{} (\omega)&=&\Sigma^{R}_{1} (\textbf{}^{}\omega) + \Sigma^{R}_{2} (\textbf{}^{}\omega).\\
\Sigma^{R}_{1} (\textbf{}^{}\omega)&=&U <n_{-\sigma, \bar{\delta}}> + 0\times (1 - < n_{-\sigma,\bar{\delta}}> )\nonumber\\
+~U^2 &\times& \frac{<n_{-\sigma, \bar{\delta}}> (1 - <n_{-\sigma, \bar{\delta}}>)} {\omega + \mu^{'} + 0 - U (1 - <n_{-\sigma, \bar{\delta}}>)}.\\
\mu^{'}&=&0.\\
\Delta_{J} &=&  \nu^{} \Delta_{J_{0}}.\\
%\end{eqnarray}
%G^{R}_{2} (\textbf{}^{}\omega)_{} &=&\frac{1-<n_{-\sigma, \Delta_{J}}>} {\omega+\mu^{'}-\Delta_{J} } + \frac{<n_{-\sigma, \Delta_{J}}>} {\omega+\mu^{'}-U}.\\
%\mu^{'}&=&0.\\
%~<n_{-\sigma, \bar{\Delta_{J}}}>&+&<n_{-\sigma, \Delta_{J}}> = \nu^{}. \\
%\Delta_{J} &=&  \nu^{} \Delta_{J_{0}} .
%\begin{eqnarray} \label{eq:eq2}
< n_{-\sigma,~x} > &=&\nu^{} f(x)^{}.~~~~~~~~~\\
< n_{-\sigma,~\bar{x}} > &=&  \nu^{} (1- f(x))^{}.~~\\
< n_{-\sigma,~x} > &+& < n_{-\sigma,~\bar{x}}>= \nu^{}.\\
%<n_{-\sigma, \bar{\delta}}> + <n_{-\sigma, \delta}> &=& \nu^{} .\\
f(x)&=&\frac{1} {1+e^{(\beta x )}}.
\end{eqnarray}
\end{subequations}
\normalsize
%%%%%%%
where as the thermodynamic parameters, $\mu$ ($\mu{'}$ and $\mu^{''}$) is the chemical potential;  $\nu$ is the occupation ratio, which is equal to "1 minus the hole concentration" and has a range of \begin{math} 0.7 < \nu < 1 \end{math} in this under-doped study;   $f(x)$ is the modified Fermi-Dirac distribution function, in which $x$ itself represents the energy difference from the chemical potential (Fermi-level);  \begin{math} \beta = 1 / k_{} T \end{math};  \begin{math} < n_{-\sigma,x} > \end{math} (weighting factor) is the thermal average of the density of opposite-spin particles as the nearest-neighbors. The number of sites per energy-level is 1 and double occupancy is not allowed, so that the combination of \begin{math} < n_{-\sigma,\bar{x}} > \end{math} and \begin{math} < n_{-\sigma,x} > \end{math} satisfies \begin{math} < n_{-\sigma,\bar{x}}> + < n_{-\sigma,x}>=1 \times \nu \end{math} at finite temperature. 
% In addition, $\Delta_{J_{}}$ is also thought to vary with temperature
%\begin{math} \nu t _{0} = 0.4 eV \end{math} and \begin{math} \nu U_{0}  = 3.2 eV \end{math} at  \begin{math} \nu = 0.8 \end{math}
%\with \begin{math} t = 0.4 \end{math} and  \begin{math} U = 3.2 \end{math} 

The $\Delta_{J}$ is the low-energy excitation per electron, and the \begin{math} \Delta_{J_{0}} \end{math} ( \begin{math} = 0.125 eV \end{math} ) is the value at \begin{math} \nu = 1.0 \end{math} and \begin{math} T = 0 K \end{math}, assuming that \begin{math} \nu  \Delta_{J_{0}} = 0.1 eV \end{math} at \begin{math} \nu = 0.8 \end{math}, where \begin{math} J = 4t^{2}/U = 0.2 eV \end{math} per two electrons.

The coefficient associated with \begin{math} \Delta_{J_{}} \end{math} is determined by the antiferromagnetic correlation ratio of the system. More specifically, in a magnetic structure with magnetic domains, appeared in such as HTSC, it is somewhat trivial that the antiferromagnetic correlation ratio lies between $\nu$ and $\nu^2$, the latter of which is expected for a system with a random distribution of holes.  

The $\delta$ is a quasi-particle pole produced from Eqs.(6) and is self-consistently determined through Eqs.(3) and Eqs.(6) as these equations describe the system in an equilibrium state.
 
The chemical potential in the hole-doped region (\begin{math} \nu < 1 \end{math}) lies at the top of the lower-band (valence band) at this stage. Note that the chemical potential in the half-filled system (\begin{math} \nu = 1 \end{math}) rises to the center of the energy-gap, which is a characteristic feature of the Mott system.
%and \begin{math} \textless n_{-\sigma}\textgreater= 1/2 \end{math} because of an electron-hole symmetry, 
% between $0$ and $\Delta$ (\begin{math} \mu=\Delta/2 \end{math})

In this model, a free electron described by these Green's functions is assumed to be subject to the same restrictions as those on a doped CuO$_{2}$ plane, hopping from one of the nearest-neighbor sites under the antiferromagnetic condition to either a vacancy site (\begin{math} 1- < n_{-\sigma, \bar{x}} > \end{math}  or \begin{math} 1- < n_{-\sigma, x} > \end{math}) or an opposite-spin site (\begin{math} < n_{-\sigma, \bar{x}}> \end{math} or \begin{math} < n_{-\sigma, x} > \end{math}) which represents a singly occupied state at the energy-level $0$ or $\Delta_{J}$. When it is added to an opposite-spin site and then hops back, the energy of this system momentarily increases to $U$.
% It is assumed that hopping into $\Delta_{J}$ in the $G^{R}_{2}(\textbf{}^{}\omega)_{}$ is possible as a quasiparticle excitation unless there is thermal excitation, so that when the temperature rises and another electron on the ground state is excited to that level, the hopping energy increases by U.
%Note that, ($1-\textless n_{-\sigma, \bar{\Delta_{J}}}>$) of $G^{R}_{1}(\textbf{}^{}\omega)_{}$ indicates the density of the empty site, while ($1-\textless n_{-\sigma, \Delta_{J}} \textgreater$) of $G^{R}_{2}(\textbf{}^{}\omega)_{}$ indicates that of the occupied state, as well as $G^{R}_{2}(\textbf{}^{}\omega)_{}$ shows the excitation of  $\Delta_{J}$ is always "1" at \begin{math} T=0 \end{math}, but when the temperature rises and the electron of the ground state is excited to that level, the excitation energy increases by U. 
%Also, the first term ($1-\textless n_{-\sigma, \Delta_{J} }\textgreater$) of Equation.(2c) assumes the existence of a quasiparticle excitation even at \begin{math} T=0 \end{math}.
%%%%%%%%%%%%%%%%%%%%%%%%%%%%%%%%%%%%%%%%%%%%%%%%%%%%%%%%%%%%%%%%%%%%%%%%%%%%%%%%%%

%%%%%%
% Self-energy
%%%%%%%%%%%%%%%%%%%%%%%%%%%%%%%%%%%%%%%%%%%%%%%%%%%%%%%%%%%%%%%%%%%%%%%%%%
\subsection{Green's function and self-energy at \begin{math} \textbf{\textit{k}} = (0,\pi)/ (\pi, 0) \end{math}}
The single-particle Green's function ($G^{R}_{2} (\omega)_{}$) and its self-energy ($\Sigma^{R}_{2} (\textbf{}^{}\omega)$) at \begin{math} (k_{x}, k_{y}) = (0, \pi)/ (\pi, 0) \end{math} in the atomic limit derived through the Dyson equation are shown as follows, at which an excitation $\Delta_{J}$ is observed.
%%%%%%%
\footnotesize
\begin{subequations}
\begin{eqnarray} \label{eq:eq4}
%\Sigma^{R}_{} (\omega)&=&\Sigma^{R}_{1} (\textbf{}^{}\omega) + \Sigma^{R}_{2} (\textbf{}^{}\omega).\\
%\Sigma^{R}_{1} (\textbf{}^{}\omega)&=&U <n_{-\sigma, \bar{\delta}}> + 0\times (1 - < n_{-\sigma,\bar{\delta}}> )\nonumber\\
%&+&U^2 \times \frac{<n_{-\sigma, \bar{\delta}}> (1 - <n_{-\sigma, \bar{\delta}}>)} {\omega + \mu^{'} + 0 - U (1 - <n_{-\sigma, \bar{\delta}}>)}.\\
G^{R}_{2} (\textbf{}^{}\omega)_{} &=&\frac{1-<n_{-\sigma, \Delta_{J}}>} {\omega+\mu^{'}-\Delta_{J} } + \frac{<n_{-\sigma, \Delta_{J}}>} {\omega+\mu^{'}-U}.\\
\Sigma^{R}_{2} (\textbf{}^{}\omega)&=&U < n_{-\sigma, \delta}> + \Delta_{J} (1 - < n_{-\sigma, \delta}> )\\
%&+&(U-\Delta)^{2} \nonumber\\
+~(U-\Delta_{J} )^{2}&\times&\frac{<n_{-\sigma, \delta}> (1-<n_{-\sigma, \delta}>)} {\omega+\mu^{'} - \Delta_{J} <n_{-\sigma, \delta}> - U(1 - <n_{-\sigma, \delta}>)}.\nonumber\\
\mu^{'}&=&0.
%\Delta_{J} &=&  \nu^{} \Delta_{J_{0}}.
%\end{eqnarray}
%\begin{eqnarray} \label{eq:eq4}
%~<n_{-\sigma, \bar{\Delta_{J}}}> + <n_{-\sigma, \Delta_{J}}> &=& \nu^{}. \\
%<n_{-\sigma, \bar{\delta}}> + <n_{-\sigma, \delta}> &=& \nu^{} .
\end{eqnarray}
\end{subequations}
\normalsize
%%%%%%
In fact, $\Sigma^{R}_{2} (\textbf{}^{} \omega)$ is an approximately positive constant \begin{math} \Delta_{J_{}} \end{math}, which will be shown in Section 5 to be the antiferromagnetic interaction energy per electron between the nearest neighbor sites of the system. If the self-energy in Eq.(3b) is such an inter-site energy, its first term due to $U$ and therefore its third term which is a second-order effect of the system will not appear in the real system, and further $\Delta_{J_{}}$ itself here already includes the effect of a second order perturbation due to $U$. In this paper, both terms are calculated because they are significantly smaller than $\Delta_{J_{}}$ at even finite temperature.

\subsection{Green's function and self-energy at the both points}
The Green's functions and the self energies per electron both at \begin{math} \textbf{\textit{k}}_{F} \end{math} and at \begin{math} \textbf{\textit{k}} = (0,\pi)/ (\pi, 0)\end{math} discussed here are denoted as $G^{R}_{} (\textbf{}^{}\omega)_{}$ and $\Sigma^{R}_{} (\textbf{}^{}\omega)$. $G^{R}_{} (\textbf{}^{}\omega)_{}$ in Eq.(4a) shows the poles of the energy structure of this system before incorporating the self-energy and is used for comparison with Eq.(6a).
%%%%%%
 \footnotesize
\begin{subequations}
\begin{eqnarray} \label{eq:eq5}
G^{R}_{} (\textbf{}^{}\omega)_{} = G^{R}_{1} (\textbf{}^{}\omega)_{} + G^{R}_{2} (\textbf{}^{}\omega)_{} .\\
\Sigma^{R}_{} (\textbf{}^{}\omega) = \Sigma^{R}_{1} (\textbf{}^{}\omega) + \Sigma^{R}_{2} (\textbf{}^{}\omega) .
\end{eqnarray}
\end{subequations}
\normalsize
%%%%%%%%%%%%%%%%%%%%%%%%%%%%%%%%%%%%%%%%%%%%%%%%%%%%%%%%%%%%%%%%%%%%%%%%%%%%%%%%%%

%%%%%%
 % Hubbard 1 approximation %%%%%%%%%%%%%%%%%%%%%%%%%%%%%%%%%%%%%%%%%%%%%%%%%%%%%%%%%%%%%%%%%%%%%
 %%%%%%
\subsection{Green's function for the Hubbard-1 approximation}
Generally, the Hubbard-1 approximation ($G^{R}_{H1} (\mathbf{\textit{k}}, \omega)_{}$) is shown as below, and estimates the Green's function poles of the system using the self-energy in the atomic limit.
%%%%%%%Correspondingly,
\footnotesize
\begin{subequations}
\begin{eqnarray} \label{eq:eq5}
G^{R}_{H1} (\mathbf{\textit{k}}, \omega)_{}&=&\frac{1 - < n_{-\sigma, \bar{\delta}}>} {\omega + \mu^{''} - \epsilon(\textbf{\textit{k}}) - \Sigma^{R}_{H1}(\omega)}. \\
\mu^{''}&=&\Sigma^{R}_{H1}(\omega=0).
%G^{R}_{H1}(\omega)_{}&=&\frac{1}{\omega+\mu^{''}-0-\Sigma^{R}_{}}+\frac{1}{\omega+\mu^{''}-\Delta-\Sigma^{R}_{}}.\\
%G^{R}_{H1}(\omega)_{}&=&\frac{\textless n_{-\sigma, -\delta/2}\textgreater}{\omega+\mu^{''}-0-\Sigma^{R}_{}}+\frac{1-\textless n_{-\sigma, -\delta/2}\textgreater}{\omega+\mu^{''}-\Delta-\Sigma^{R}_{}}.\\
%&+&\frac{\textless n_{-\sigma, -\delta/2}\textgreater+\textless n_{-\sigma, +\delta/2}\textgreater}{\omega+\mu^{''}-U-\Sigma^{R}_{}}.\\
%\mu^{''}&=&\Sigma^{R}_{1}(\omega=0).\\
\end{eqnarray}
\end{subequations}
\normalsize
%%%%%%
where $\epsilon(\textbf{\textit{k}})$ represents the dispersion relation of the non-interacting energy across the entire $\textbf{\textit{k}}$, and $\Sigma^{R}_{H1}(\omega)$ is the local self-energy of the system in the atomic limit. 

In this study, the Green's function for the Hubbard-1 approximation is described in Eqs.(6). $G^{R}_{H1}(\omega)_{}$ combines the Green's function at \begin{math} \textbf{\textit{k}} = \textbf{\textit{k}}_{F} \end{math} and that at \begin{math} \textbf{\textit{k}} = (0,\pi)/ (\pi, 0) \end{math} in the second term, in terms of expressing the energy structure of the system incorporating self-energy.
%(\begin{math} \epsilon(\textbf{\textit{k}}_{F}) = 0 $G^{R}_{H1}(\omega)_{}$
%%%%%%
\footnotesize
\begin{subequations}
\begin{eqnarray} \label{eq:eq6}
G^{R}_{H1}(\omega)_{}&=&\frac{1 - < n_{-\sigma, \bar{\delta}}>} {\omega + \mu^{''} - 0 - \Sigma^{R}_{1}(\omega)} \nonumber\\
&+&\frac{1 - < n_{-\sigma, \delta}>} {\omega+\mu^{'' }- \Delta_{J} - (\Sigma^{R}_{1}+ \Sigma^{R}_{2})(\omega)}.\\
\mu^{''}&=&\Sigma^{R}_{1}(\omega=0).
%\Delta_{J} &=&  \nu^{} \Delta_{J_{0}}.
%\end{eqnarray}
%\begin{eqnarray} \label{eq:eq6-2}
%< n_{-\sigma, \bar{\delta}}> + < n_{-\sigma, \delta}>~=~\nu^{} .
\end{eqnarray}
\end{subequations}
\normalsize
%%%%%%
The first term indicates a pole at the Fermi-level and the second term creates a pole ($\delta$) that is shifted from the original excitation ($\Delta_{J}$) above the Fermi-level at \begin{math} \textbf{\textit{k}} = (0,\pi)/ (\pi, 0) \end{math}. The chemical potential is set to \begin{math} \mu^{''} = \Sigma^{R}_{1} (\omega = 0) \end{math}. The additional self-energy $\Sigma^{R}_{2}(\textbf{}^{} \omega)$ is assumed to overlap the local self-energy (on-site) $\Sigma^{R}_{1}(\textbf{}^{} \omega)$ at \begin{math} \textbf{\textit{k}} = (0,\pi)/ (\pi, 0) \end{math}.
%the local self-energy $\Sigma^{R}_{1}(\textbf{}^{} \omega)$ at $[0,\pi]$ and $[\pi, 0]$
%$\textless n_{-\sigma,x}\textgreater$ of the lower-band is set to $\textless n_{-\sigma,\bar{\delta}}\textgreater$ and that of the upper-band is set to $\textless n_{-\sigma,\delta}\textgreater$.
%%%%%%%%%%%%%%%%%%%%%%%%%%%%%%%%%%%%%%%%%%%%%%%%%%%%%%%%%%%%%%%%%%%%%%%%%%%%%%%%%%

%%%%%%%%%%%%%%%%%%%%%%%%%%%%%%%%%%%%%%%%%%%%%%%%%%%%%%%%%%%%%%%%%%%%%%%%%%%%%%%%%%%%%
% Results %%%%%%%%%%%%%%%%%%%%%%%%%%%%%%%%%%%%%%%%%%%%%%%%%%%%%%%%%%%%%%%%%%%%%%%%%%%%%%
%%%%%%%%%%%%%%%%%%%%%%%%%%%%%%%%%%%%%%%%%%%%%%%%%%%%%%%%%%%%%%%%%%%%%%%%%%%%%%%%%%%%%
\section{Results}

\subsection{Green's function}
The study of the electron excitation structures in the Mott system through the Hubbard-1 approximation in this study shows the following results. In the Figures, the Green's function $G^{R}(\omega)$ (Gray lines) excludes the self-energy as shown in Eq.(4a) and $G^{R}_{H1}(\omega)_{}$ (Red lines) includes the self-energy as shown in Eq.(6a), and both focus on the Fermi-level (\begin{math} \epsilon(\textbf{\textit{k}}_{F}) = 0 \end{math}) and the low-level excitation at \begin{math} \textbf{\textit{k}} = (0,\pi)/ (\pi, 0) \end{math}. The self-energy effect in  $G^{R}_{H1}(\omega)_{}$ shifts an original excitation $\Delta_{J}$ to a new excitation $\delta$ of the system. 
%In the Figures, $G$($\omega$) represents the poles of Green's functions, both $G^{R}(\omega)$ and $G^{R}_{H1}(\omega)_{}$.

Setting the original excitation as \begin{math} \Delta_{J} = 0.1eV \end{math} results in a shifted excitation \begin{math} \delta = 0.038eV \end{math} as shown in Fig(Figure).1, where \begin{math} \nu = 0.8 \end{math}, \begin{math} T=0 K \end{math}, \begin{math} U = 3.2eV \end{math}, and \begin{math} U/t = 8 \end{math}. The shifted excitations ($\delta$) for various original excitations ($\Delta_{J}$) near \begin{math} \Delta_{J} = 0.1eV \end{math} are shown in Fig.2, with the same values for $\nu$, $T$, and $U/t$ as above : ($\Delta_{J}$[$eV$], $\delta$[$eV$]) = (0.075, 0.029), (0.1, 0.038), and (0.125, 0.047).

Fig.1 and Figs.3-5 show the \begin{math} \omega - G(\omega) \end{math} (Green's function) curves, which indicate the poles of the shifted excitations ($\delta$) at various occupation ratios ($\nu$) and temperatures ($T$) : ($\nu$, $T$[$K$], $\delta$[$eV$]) = (0.8, 0, 0.038), (0.9, 0, 0.021), (0.7, 0, 0.051), and (0.8, 100, 0.040) respectively. 

Besides, Fig.6 shows the wider range of the Green's function curves for Fig.1 with \begin{math} \nu = 0.8 \end{math}, \begin{math} T = 0K \end{math}, \begin{math} \Delta_{J} = 0.1eV \end{math} and \begin{math} U/t = 8 \end{math}. In this figure, the poles of $G^{R}_{H1}(\omega)$ appear at 0, 0.038, 3.2, and 3.36~$eV$, while the poles of $G^{R}_{}(\omega)$ appear at 0, 0.1(J per electron), 3.2~$eV$. Fig.7 shows the magnitude of the shifted excitation ($\delta$) per electron as a function of the doping-ratio (1-$\nu$).
%%%%%%%

%%%%%%%
% Hubbard 1 approximation %%%%%%%%%%%%%%%%%%%%%%%%%%%%%%%%%%%%%%%%%%%%%%%%%%%%%%%%%%%%%%%%%%%%%
% Figures 1
%%%%%%
\begin{figure}
\includegraphics[height=5.0 cm, bb=0 0 846 594]{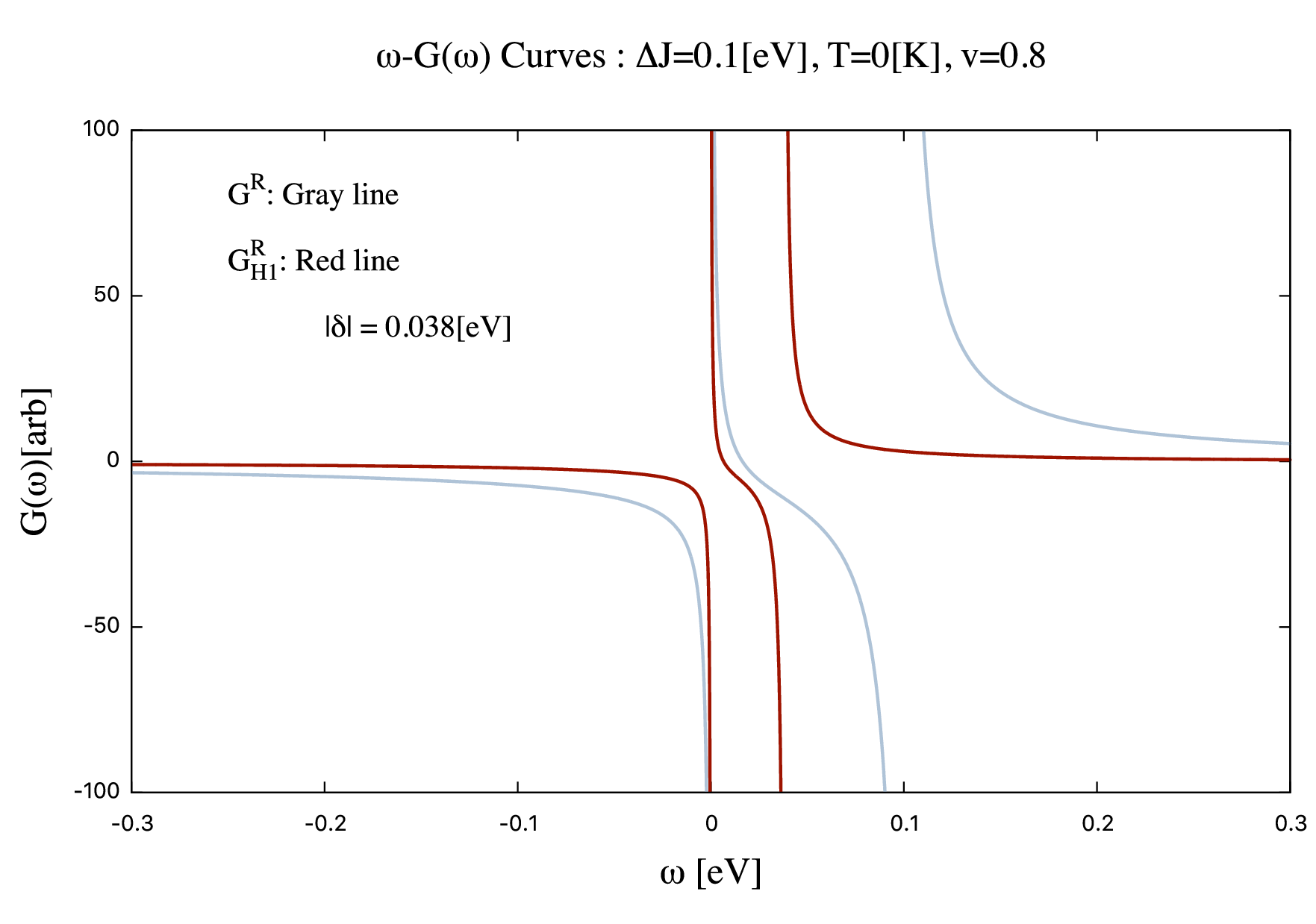}
\caption{$\omega$-$G$($\omega$) curves for a single particle Green's function: Gray ($G^{R} (\omega)$) denotes the Green's function without self-energy and Red ($G^{R}_{H1}(\omega)$) denotes that including self-energy. $|\delta| \sim 0.038 eV$ ($2\delta \sim 0.076 eV$), when $\nu = 0.8$, $T = 0 K$, $\Delta_{J} = 0.1 eV$, and $U/t = 8$.}
\label{f1}
\end{figure}

% Figures 2
\begin{figure}
\includegraphics[height=5.0 cm, bb=0 0 846 594]{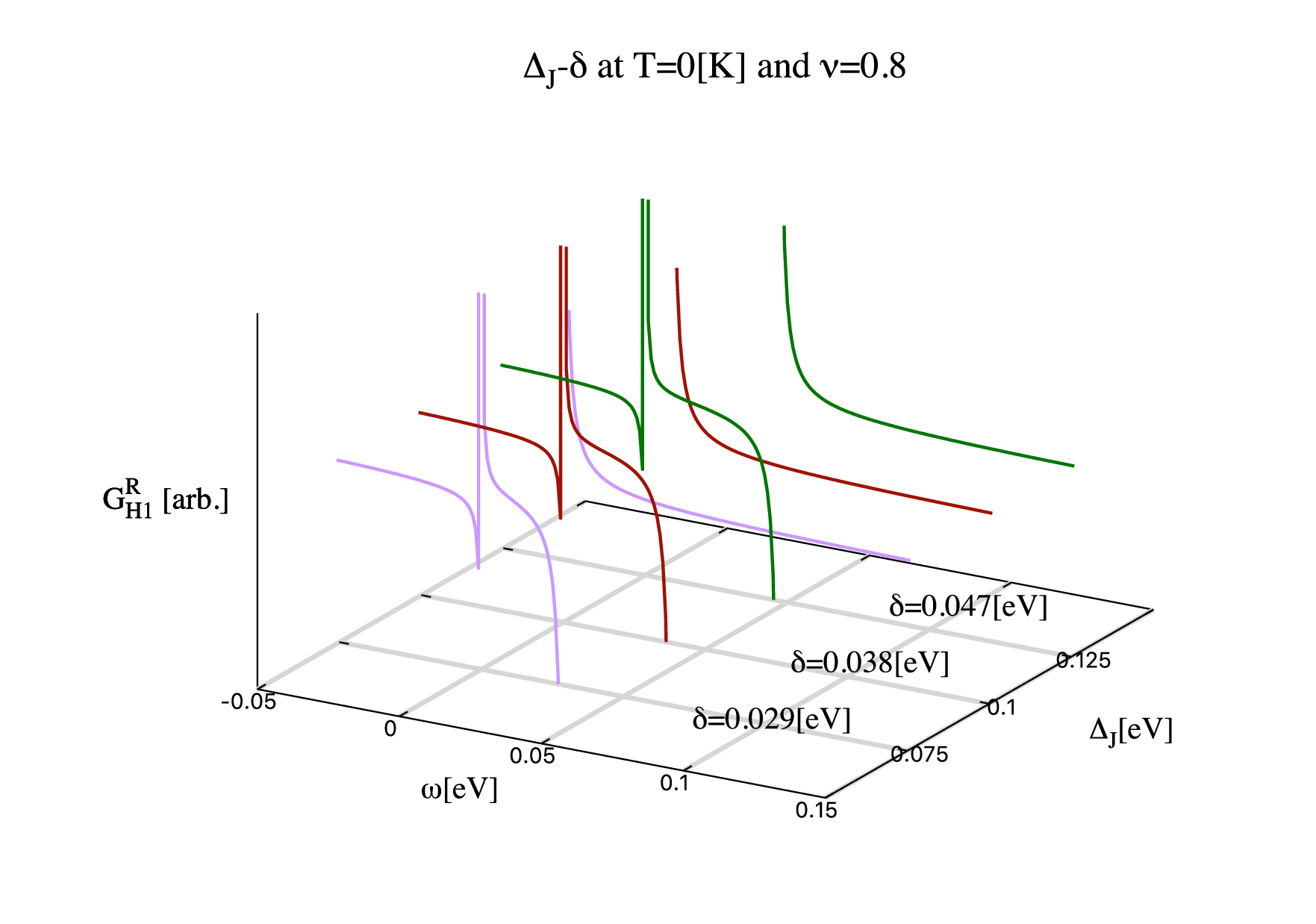}
\caption{$\Delta_{J} - \delta$: ($\Delta_{J}$ [$eV$], $\delta$ [$eV$]) = (0.075, 0.029), (0.1, 0.038), and (0.125, 0.047), where $\nu = 0.8$, $T = 0 K$, and $ U/t = 8$.}
\label{f2}
\end{figure}
%gcc /Users/tanakakeishichirou/Desktop/Programs/APS_2021S/Calc_2SH2_GreenFunction_0615_3_J_Gap_Curve.c
%load "/Users/tanakakeishichirou/Desktop/Programs/APS_2021A/gnuplot_2SH_J_delta_2021S_0817.txt"
%The original $\Delta_{J}$ yields the resulting $ \delta $ per electron through the self-energy effect as ,

% Figures 3
\begin{figure}
\includegraphics[height=5.0 cm, bb=0 0 846 594]{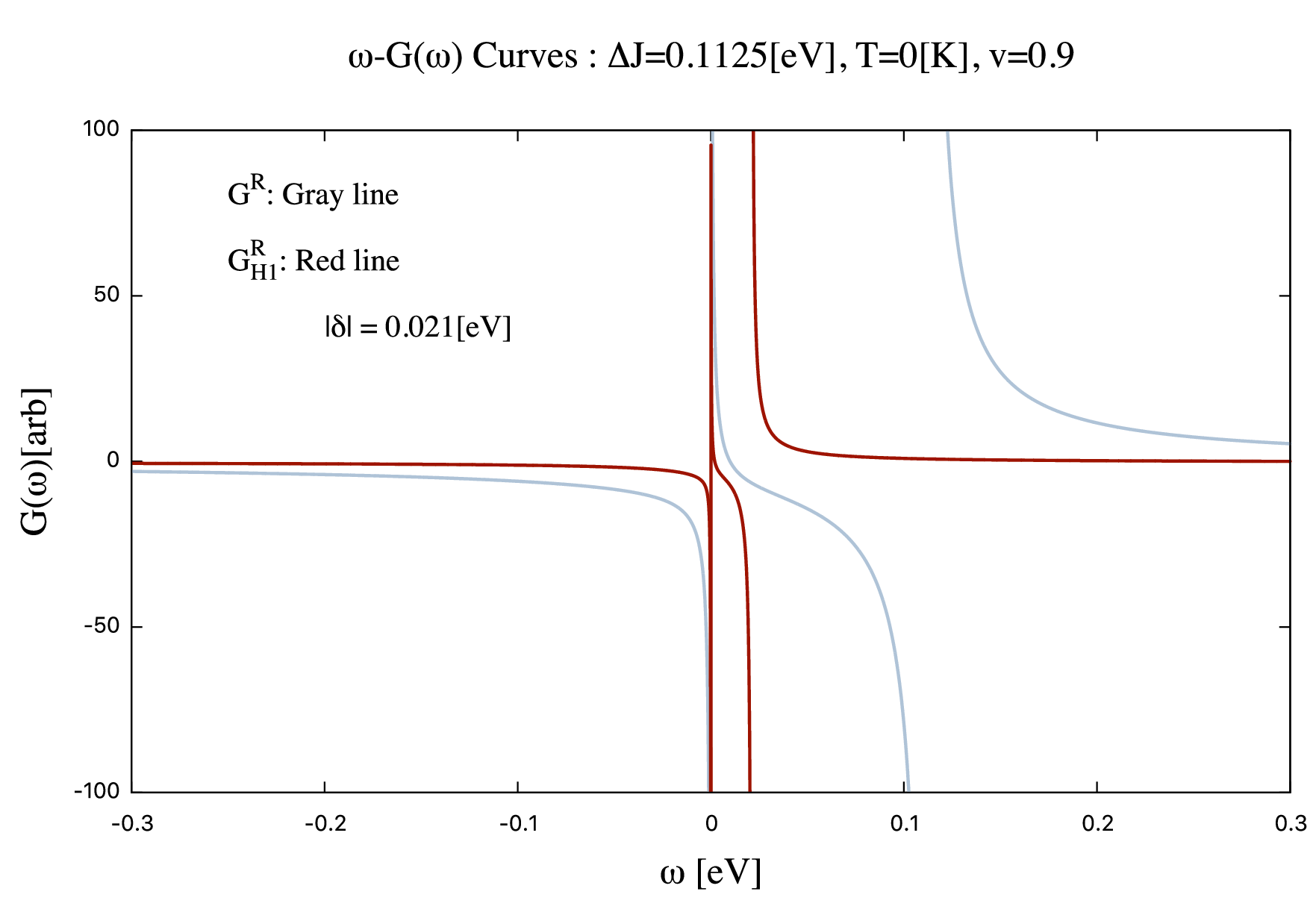}
\caption{$\omega$-$G$($\omega$) curves : $|\delta| \sim 0.021 eV$, when $\nu = 0.9$, $T = 0 K$, and $\Delta_{J} = 0.1125 eV$. Refer to Fig.1.}
\label{f3}
\end{figure}

% Figures 4
\begin{figure}
\includegraphics[height=5.0 cm, bb=0 0 846 594]{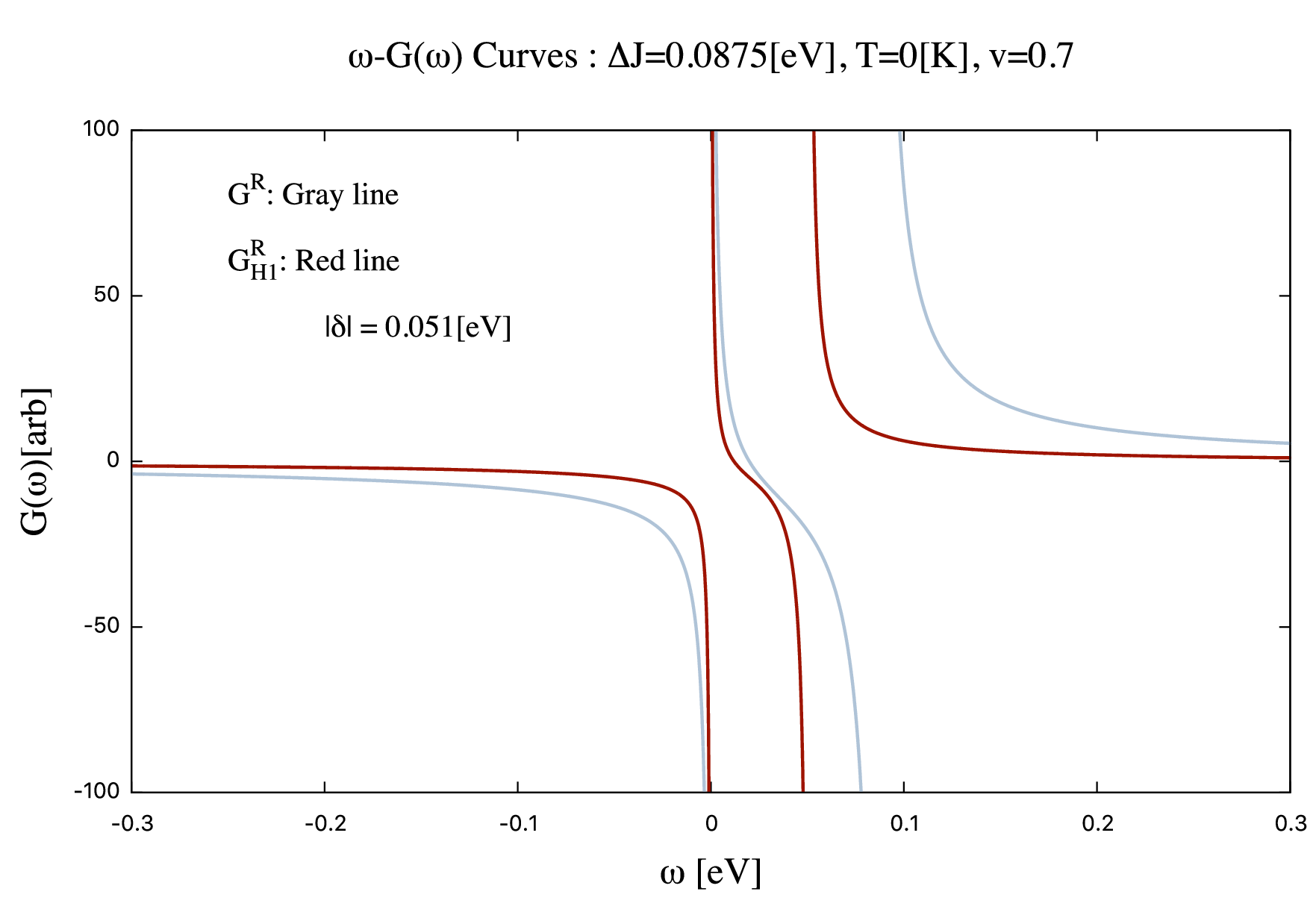}
\caption{$\omega$-$G$($\omega$) curves for a single particle: $|\delta| \sim 0.051 eV$, when $\nu = 0.7$, $T = 0 K$, and $\Delta_{J} = 0.0875 eV$. Refer to Fig.1.}
\label{f4}
\end{figure}

% Figures 5
\begin{figure}
\includegraphics[height=5.0 cm, bb=0 0 846 594]{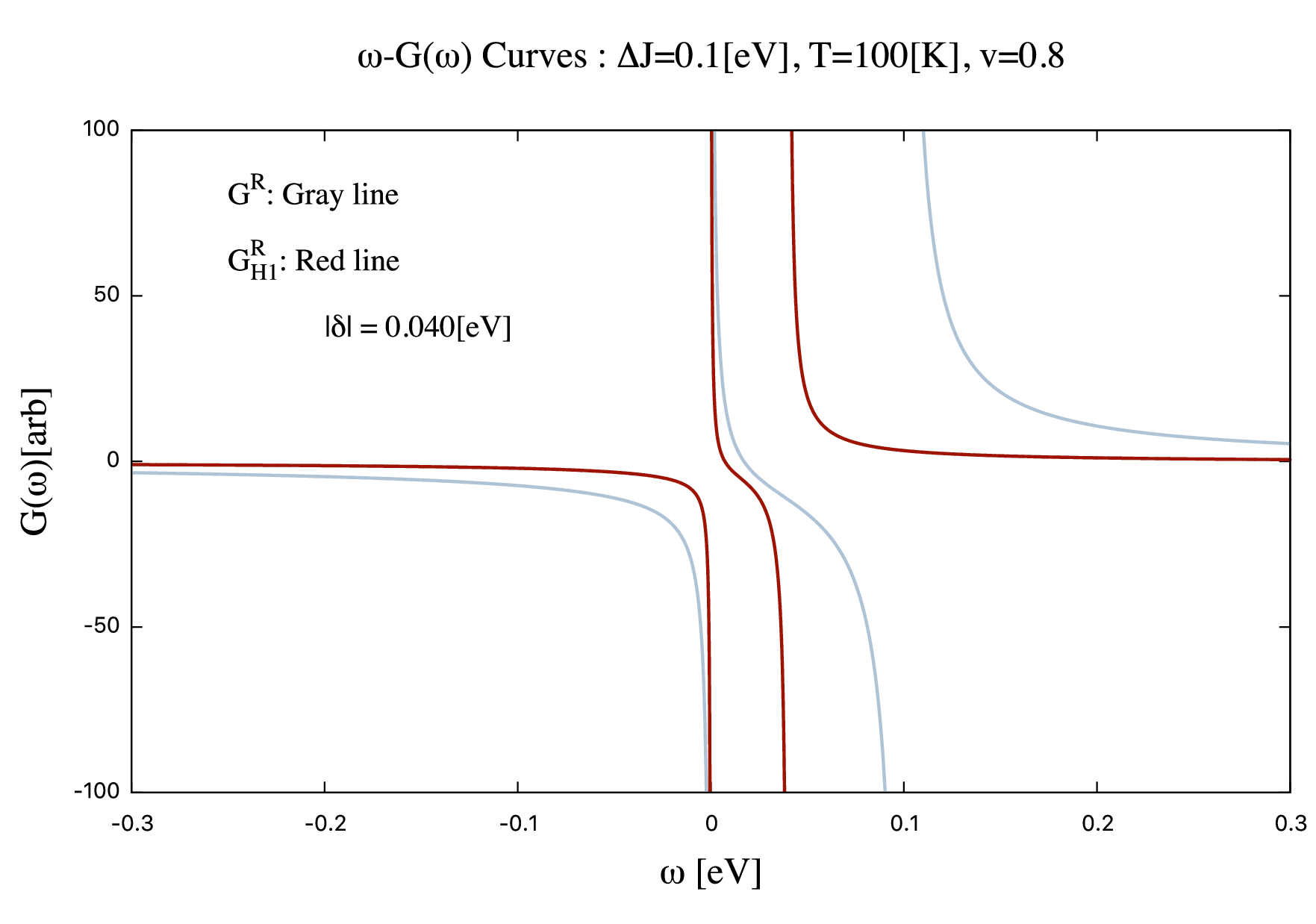}
\caption{$\omega$-$G$($\omega$) curves: $|\delta| \sim 0.040 eV$, when $\nu = 0.8$, $T = 100 K$, and $\Delta_{J} = 0.1 eV$. $\mu^{"}$ and $\Delta_{J}$ are set to be the same as those at $T = 0 K$. Refer to Fig.1.}
\label{f5}
\end{figure}

% Figures 6
\begin{figure}
\includegraphics[height=5.0 cm, bb=0 0 846 594]{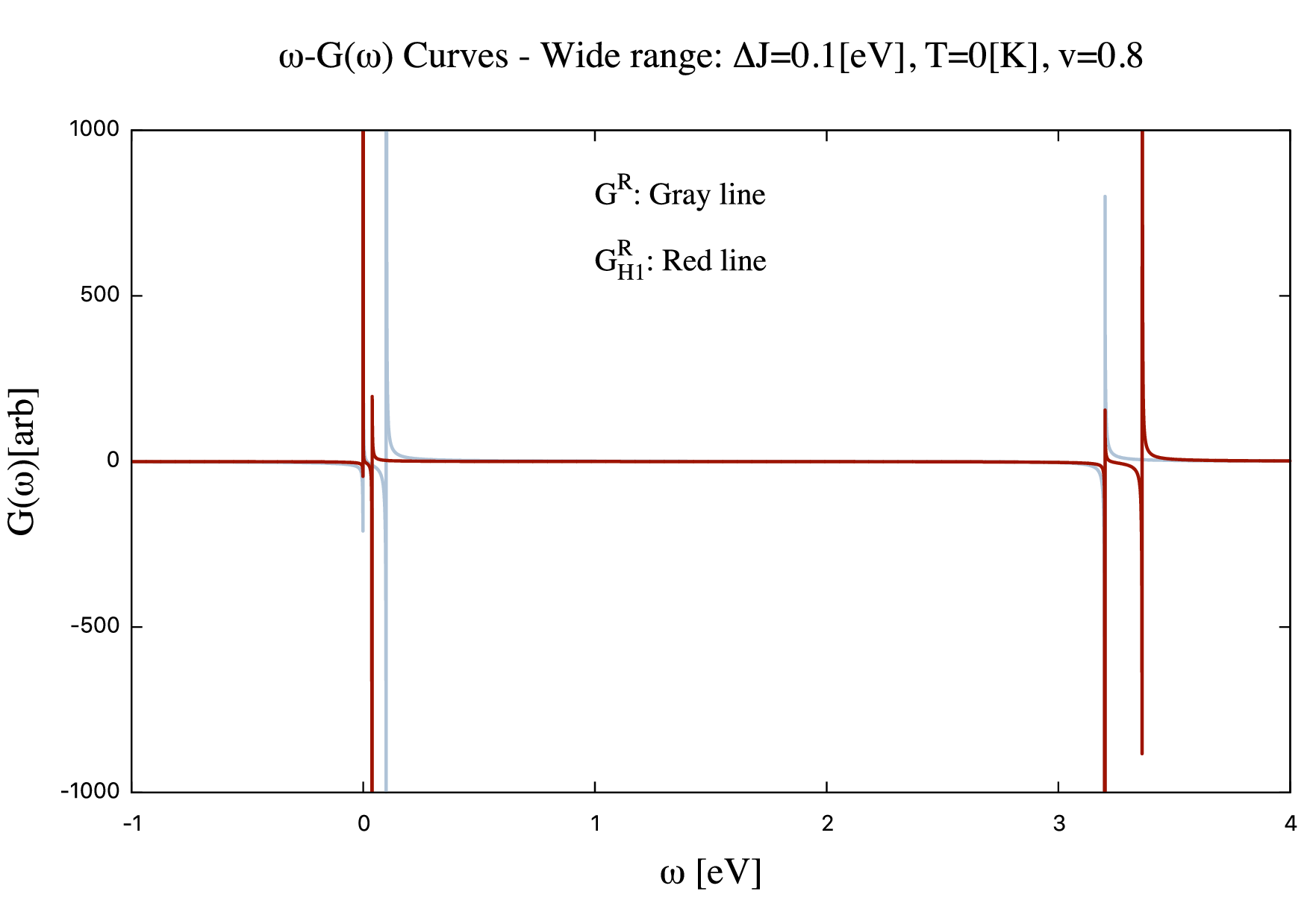}
\caption{$\omega$-$G$($\omega$) curves (Wide range):  $|\delta| \sim 0.038 eV$, when $\nu = 0.8$, $T = 0 K$, and $\Delta_{J} = 0.1 eV$. The poles of $G^{R}(\omega)$ appear at 0, 0.1, and 3.2$eV$. The poles of $G^{R}_{H1}(\omega)$ appear at 0, 0.038, 3.2, and 3.36$eV$. Refer to Fig.1.}
\label{f6}
\end{figure}
%gcc /Users/tanakakeishichirou/Desktop/Programs/APS_2021A/Calc_GW_Hubbard1_0615.c
%load "/Users/tanakakeishichirou/Desktop/Programs/APS_2021A/gnuplot_2SH_GreenFunction_2020_0425"

% Figures 7 Doping-ratio - Pseudo-gaps per electron plots
\begin{figure}
\includegraphics[height=5.0 cm, bb=0 0 846 594]{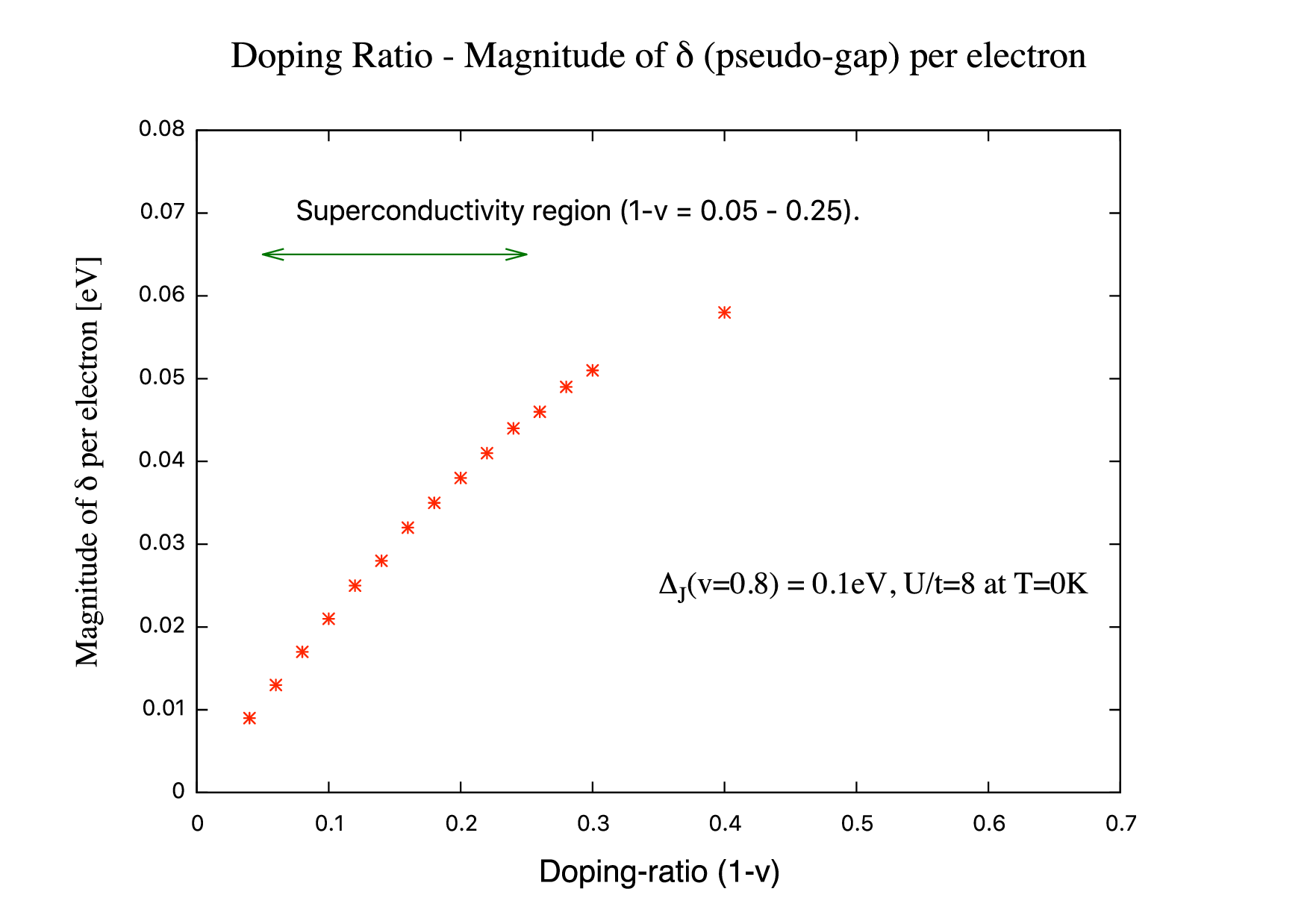}
\caption{Doping-ratio - Magnitude of $\delta$ (pseudogap) per electron, which draws a dome shape, where $T = 0 K$, $\Delta_{J} = 0.1 eV$ at $\nu = 0.8$, and $U/t = 8$.}
\label{f7}
\end{figure}
%%%%%%

%\clearpage
% Effective mass %%%%%%%%%%%%%%%%%%%%%%%%%%%%%%%%%%%%%%%%%%%%%%%%%%%%%%%%%%%%%%%%%%%%%%%%%
%%%%%%
\subsection{Effective mass}

The effective-mass ratio is expressed as \begin{math} m^{*}/m_{} = 1/Z \end{math} at \begin{math} T = 0 K \end{math} in the Fermi liquid theory\cite{reference35}, where $\textit{Z}$ is the quasi-particle renormalization factor as below. There is a discontinuity in the momentum distribution at \begin{math} \textbf{\textit{k}} = \textbf{\textit{k}}_{F} \end{math} and \begin{math} T = 0 K \end{math}.

%%%%%%
% Equation 7
%%%%%%
\footnotesize
\begin{subequations}
\begin{eqnarray} \label{eq:eq7}
m^{*}/m_{} &=& 1/Z ~~~\mbox{($T = 0K$)} . \\
\textit{Z}&=&[1 - \partial Re\Sigma^{R}_{1} (\omega)/ \partial \omega ]^{-1}|_{\textbf{\textit{k}} = \textbf{\textit{k}}_{F},\omega = 0^{+}}.
\end{eqnarray}
\end{subequations}
\normalsize
%%%%%%

The effective-mass ratio \begin{math} m^{*}/m_{} \end{math} of this model varies from 3 to 10 at \begin{math} T =  0 K \end{math} as the occupation ratio $\nu$ changes from 0.7 to 0.9, a range associated with the superconducting state in HTSCs \cite{reference7} as shown in Fig.8, where \begin{math} \Delta_{J} =  0.1 eV \end{math} at  \begin{math} \nu = 0.8 \end{math} and \begin{math} U/t=8 \end{math}: ($\nu$, $m^{*}/m_{}$)=  (0.9, 10.0), (0.8, 5.0), (0.7, 3.3), and (0.6, 2.5). The self-energy \begin{math} \Sigma^{R}_{1} (\omega) \end{math} changes from positive to negative when $\omega$ changes from negative to positive at \begin{math} \textbf{\textit{k}} = \textbf{\textit{k}}_{F} \end{math} and \begin{math} T =  0 K \end{math}. The effective-mass ratio depends on the occupation ratio $\nu$, which is also related to the original excitation $\Delta_{J}$. 
%$\Delta_{J}$=$\nu$$\Delta_{J0}$
%, to which the Fermi liquid theory can be applied,

%%%%%%
% Figures 8, Effective mass
%%%%%%
\begin{figure}
\includegraphics[height=5.0 cm, bb=0 0 846 594]{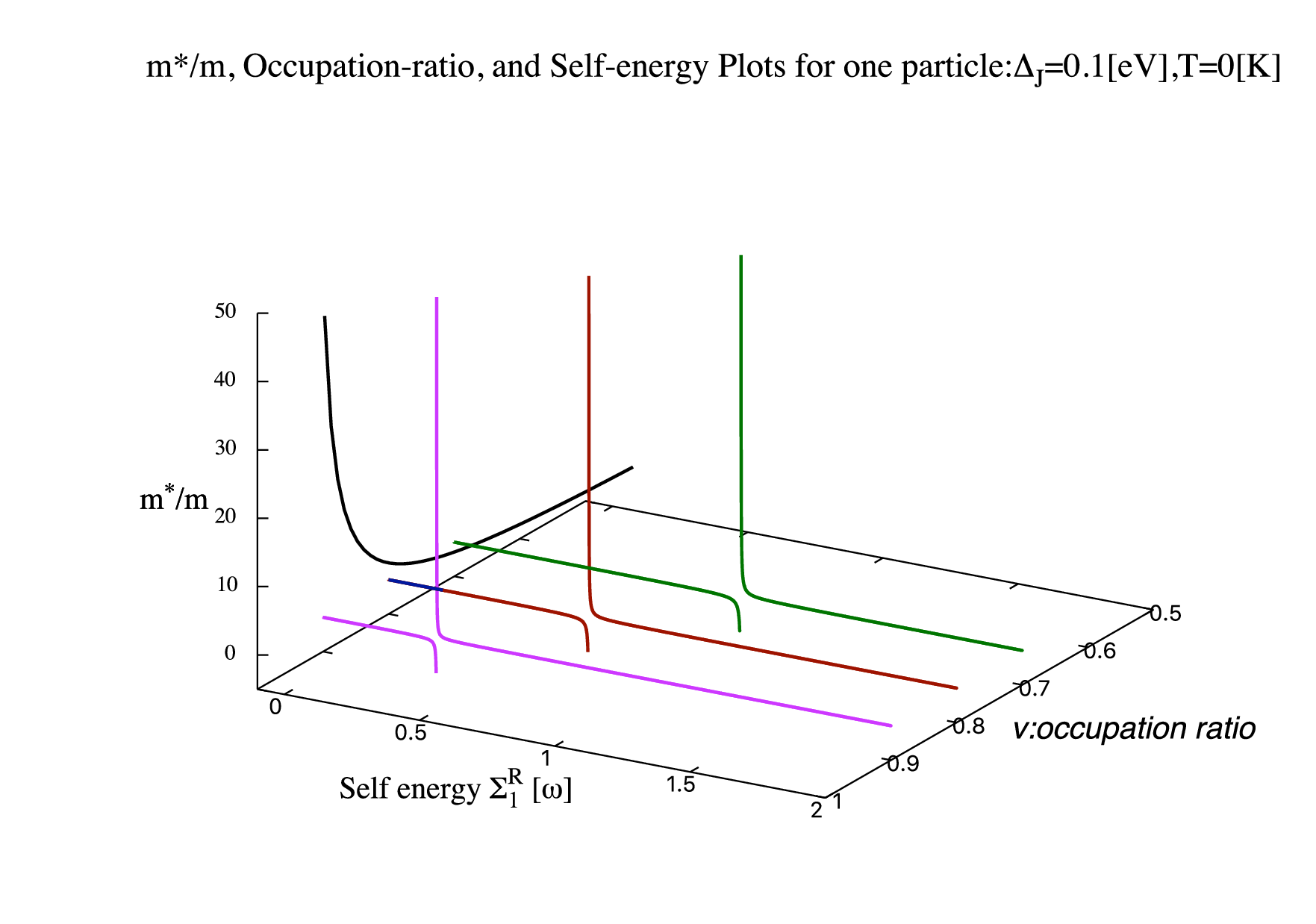}
\caption{Occupation ratio ($\nu$) - Effective-mass ratio ($m^{*}/m_{}$) : The effective-mass ratio $m^{*}/m_{}$ varies $3 - 10$ at $T = 0 K$ as the occupation ratio $\nu$ changes $0.7 - 0.9$, where $\omega = 0^{+}$, $T = 0 K$, $\Delta_{J} = 0.1 eV$ at $ \nu = 0.8 $, and $U/t = 8$. The self-energy $\Sigma^{R}_{1}$ is negative at $\omega = 0^{+}$.}
\label{f8}
\end{figure}
%gcc /Users/tanakakeishichirou/Desktop/Programs/APS_2020A/Calc_2021S_EffectiveMass_0506.c
%gcc /Users/tanakakeishichirou/Desktop/Programs/Data/2021S_EffectiveMass_0506.dat
%load "/Users/tanakakeishichirou/Desktop/Programs/APS_2020A/gnuplot_2SH_EffectiveMass_2021S_0508.txt"
%%%%%%%%%%%%%%%%%%%%%%%%%%%%%%%%%%%%%%%%%%%%%%%%%%%%%%%%%%%%%%%%%%%%%%%%%%%%%%%%%%%%%

\clearpage
%%%%%%%%%%%%%%%%%%%%%%%%%%%%%%%%%%%%%%%%%%%%%%%%%%%%%%%%%%%%%%%%%%%%%%%%%%%%%%%%%%%%%
% Discussion %%%%%%%%%%%%%%%%%%%%%%%%%%%%%%%%%%%%%%%%%%%%%%%%%%%%%%%%%%%%%%%%%%%%%%%%%%%%%%
%%%%%%%%%%%%%%%%%%%%%%%%%%%%%%%%%%%%%%%%%%%%%%%%%%%%%%%%%%%%%%%%%%%%%%%%%%%%%%%%%%%%%
\section{Discussion}

This section mainly verifies the results derived from the model presented in this study, using the past experimental results. It is assumed that Bi2212 exhibits a pseudogap of approximately 0.04eV per electron as its optimal value under an occupancy ratio \begin{math} \nu=0.8 \end{math} and \begin{math} T = 0K \end{math} \cite{reference7}.
%Now, the physical characteristic values in the optimum doping region seems most appropriate to assess the magnitude of the excitation in Bi2212 and therefore are used as the criteria for this evaluation. In this study, this region for both the presented model and the experimental results of Bi2212 is assumed to correspond to the occupancy ratio $\sim 0.8$ and the gap-size per electron $\sim 0.04~eV$ \cite{reference7}.
% where the range of the occupation ratio is \begin{math}0.7<\nu<1.0\end{math}, a superconducting region of Bi2212s \cite{reference7}. 

Firstly, when the original excitation is set as \begin{math} \Delta_{J}=0.1 eV \end{math} per electron, the resulting excitation \begin{math} \delta = 0.038 eV \end{math} per electron appears at \begin{math} \nu=0.8 \end{math} and \begin{math} T = 0 K \end{math}. Notably, the magnitude of $0.1 eV$ per electron is of the same order as the exchange interaction constant $J$ of the Mott system when \begin{math}  t =  0.4 eV\end{math} and \begin{math} U = 3.2 eV \end{math}. That is, in this condition, \begin{math} J = 4t^{2}/U = 0.2 eV \end{math}, corresponding to $2 \Delta_{J}$ at \begin{math} \nu=0.8 \end{math} per two electrons.
 %the self-energy effect of this system compresses an original spectral peak \begin{math} \Delta_{J}=0.1~eV ~ (2\Delta_{J} \sim J \sim 4t^2/U) \end{math} into the resulting spectral peak \begin{math} \delta \sim 0.038~eV \end{math} per electron at  \begin{math}T = 0K \end{math} at \begin{math} \nu=0.8 \end{math}, which is rather close to the optimal gap-size per electron of Bi2212s in an underdoped region (\begin{math} \sim 0.04~eV \end{math}) \cite{reference7}, where $J$ is the exchange interaction constant of the Mott system. 

Additionally, both the occupancy-ratio dependence and the temperature dependence of the magnitude of $\delta$ in this study closely resemble the behavior of the pseudogaps in Bi2212. The smaller the occupancy ratio is, the larger the magnitude of $\delta$ is, and as the temperature increases, $\delta$ increases slightly. As the occupation ratio approaches zero, $\delta$ converges to the original excitation $\Delta_{J}$, and when the occupation ratio reaches unity, $\delta$ vanishes. The relationship between occupancy ratios ($\nu$) and the magnitudes of shifted excitations ($\delta$, pseudogaps) draws a dome shape as illustrated in Fig.7.

Secondly, the effective mass calculated in this model aligns well with the experimental results of HTSCs \cite{reference36,reference37,reference38}. In particular, the effective-mass ratio \begin{math} m^{*}/m_{} \end{math} in this model ranges from 3 to 10 at \begin{math} T = 0K \end{math}, as the occupation ratio $\nu$ varies from 0.7 to 0.9. This is because, as the occupation ratio decreases and/or temperature increases, the pole of the self-energy in this system moves upward from the Fermi-level to $U$, and accordingly the rate of change of the self-energy at the Fermi-level decreases. 

Thirdly therefore, the estimation of the self-energy $\Sigma^{R}_{1}$ is appropriate, because the effective mass is determined from the self-energy at \begin{math} \textbf{\textit{k}}^{} = \mathbf{\textit{k}_{F}} \end{math}, \begin{math} \omega = 0 \end{math}, and \begin{math}T = 0 \end{math}, in addition to \begin{math} \Sigma^{R}_{1}(0) = 0 \end{math}. 

Fourthly, \begin{math} \Sigma^{R}_{2}(\omega) \end{math} is a positive energy due to the antiferromagnetic interaction per electron at \begin{math} \textbf{\textit{k}}^{} = (0, \pi)/(\pi, 0) \end{math} when \begin{math} T =  0 \end{math}, and presumably an inter-site local self-energy. This is because the antiferromagnetic interaction of this system is a property arising between nearest neighbor sites. That is, $\Sigma^{R}_{1}(\omega)$ is the on-site local self-energy due to $U$, and $\Sigma^{R}_{2}(\omega)$ is the inter-site local self-energy due to the antiferromagnetic interaction, both of them lie at high symmetry points.

Furthermore if \begin{math} \Sigma^{R}_{1}(\omega) + \Sigma^{R}_{2}(\omega) \end{math} is applied instead of \begin{math} \Sigma^{R}_{1}(\omega) \end{math} in the first term of Eq.(6a), the calculated results do not match the experimental results. If \begin{math} \Sigma^{R}_{1}(\omega) \end{math} is only applied as the self-energy in the second term of Eq.(6a), the excitation $\delta$ per electron that appears at \begin{math} \nu=0.8 \end{math} and \begin{math} T = 0K \end{math} is $0.02 eV$.

Note that, Fig.9 shows the gap $\delta$ in the \begin{math} \omega - G^{R}_{H1}( \omega ) \end{math} curve becomes symmetric with respect to the chemical potential, when setting \begin{math} \mu^{''} = \Sigma^{R}_{1}(\omega=0)+ \Sigma^{R}_{2}(\omega=0) \end{math} in Eq.(6b), where the $ |\delta| $ ($\sim 0.040eV$) is slightly larger than that of Fig.1 due to the difference in the self-energy imposed on the excited levels. 
%%%%%%%%%%%%%%%%%%%%%%%%%%%%%%%%%%%%%%%%%%%%%%%%%%%%%%%%%%%%%%%%%%%%%%%%%%%%%%%%%%

%%%%%%
% Figures 9
%%%%%%
\begin{figure}
\includegraphics[height=5.0 cm, bb=0 0 846 594]{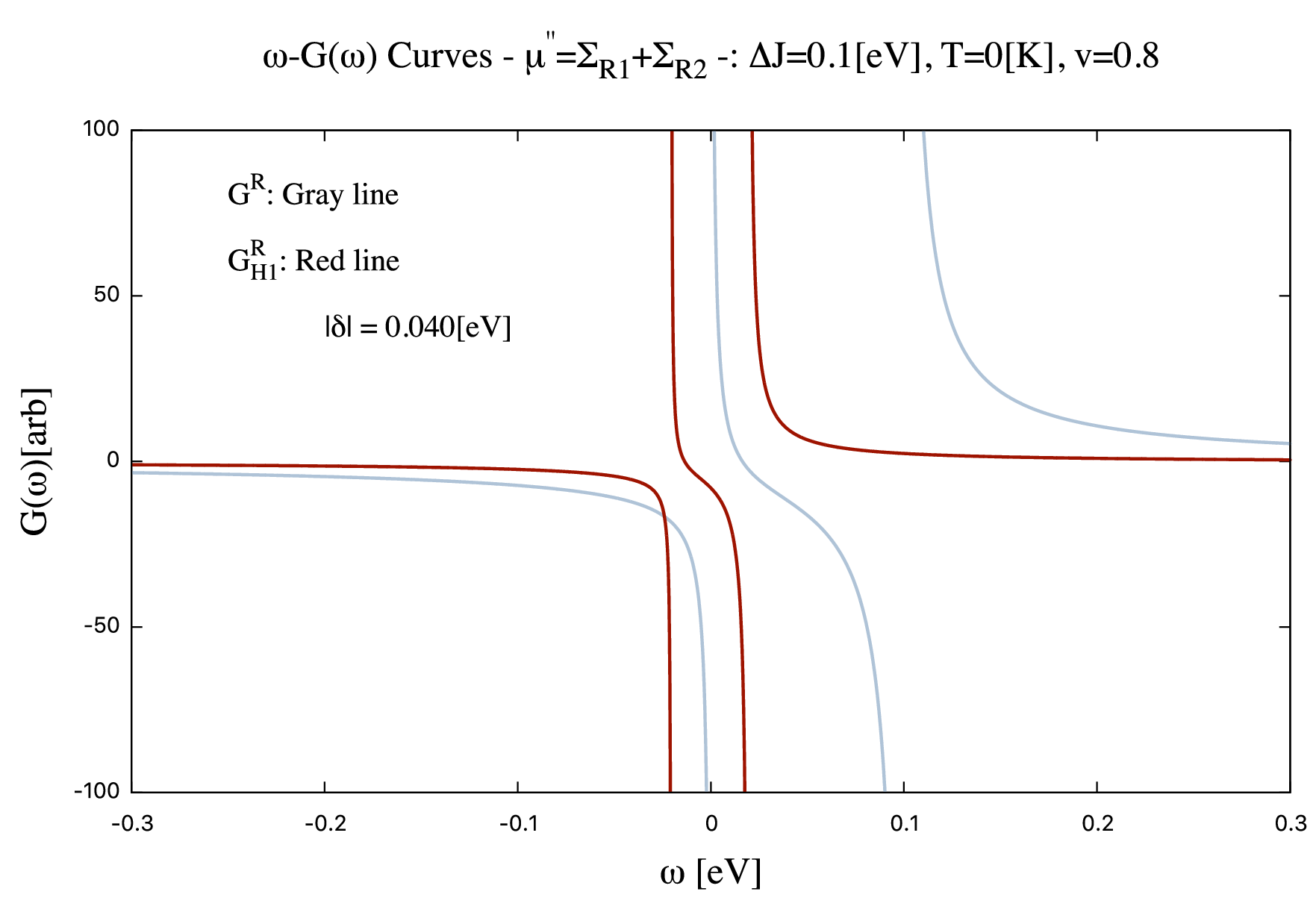}
\caption{ $\omega$-$G$($\omega$) curves: Gray ($G^{R}( \omega )$) denotes the Green's function without self-energy and Red ($G^{R}_{H1}( \omega )$) denotes that including self-energy. The chemical potential $\mu^{''} = \Sigma^{R}_{1}(\omega=0)+\Sigma^{R}_{2}(\omega=0)$. $|\delta|$ $\sim$ $0.040 eV$, when $\nu = 0.8$, $T = 0 K$, and $\Delta_{J} = 0.1 eV$. Refer to Fig.1.}
\label{f8}
\end{figure}
%%%%%%%%%%%%%%%%%%%%%%%%%%%%%%%%%%%%%%%%%%%%%%%%%%%%%%%%%%%%%%%%%%%%%%%%%%%%%%%%%%

%%%%%%%%%%%%%%%%%%%%%%%%%%%%%%%%%%%%%%%%%%%%%%%%%%%%%%%%%%%%%%%%%%%%%%%%%%%%%%%%%%
% Conclusions %%%%%%%%%%%%%%%%%%%%%%%%%%%%%%%%%%%%%%%%%%%%%%%%%%%%%%%%%%%%%%%%%%%%%%%%%%
%%%%%%%%%%%%%%%%%%%%%%%%%%%%%%%%%%%%%%%%%%%%%%%%%%%%%%%%%%%%%%%%%%%%%%%%%%%%%%%%%%
\section{Conclusions}

This study constructed the underlying system for HTSC cuprates in the under-doped region, specifically the Mott electron system with a low-energy excitation, and examined its energy excitation structure in terms of self-energy, while verifying the calculation of self-energy with the effective mass ratio. This model incorporates an additional self-energy at another high symmetry point, \begin{math} \textbf{\textit{k}}^{} = (0, \pi)~and~ (\pi, 0) \end{math}, in addition to the self-energy that is independent of $\textbf{\textit{k}}$.

The results suggest the pseudogap in HTSC is attributed to the self-energy effect on a quasi-particle excitation of the Mott system, and the energy scale of this excitation is $J$ (\begin{math} \sim 4t^{2}/U \end{math}) per two electrons. 

It is quite possible that the excitation of \begin{math} 4t^{2}/U \end{math} appears in this system, since \begin{math} 4t^{2}/U \end{math} is the energy of a quadruple degenerate second-order perturbation in the Mott system, which is the origin of its magnetism. Furthermore, the self-energy estimate appears to be adequate with respect to its effective mass ratio, which is calculated from the self-energy at \begin{math} \omega = 0 \end{math}.
%which is the offset (\begin{math} \Sigma^{R}_{1} (\textbf{}^{} 0) = 0 \end{math}) and the rate of change at \begin{math} \omega = 0 \end{math}. 

Concerning the effective-mass ratio, it also suggests the underlying system of HTSC evolves from an insulator, through the suppression of self-energy, to the Fermi liquid in the superconducting region of HTSC.

It follows from the above that, the underlying system of HTSC is depicted as the Fermi liquid state with a low-energy excitation, that activates antiferromagnetic correlations while being enveloped in a cloud of self-energy, both of which, that is, antiferromagnetic correlations and self-energy are inherent in the Mott system in the under-doped region. 

At last, the possibility of the equivalence between the pseudogap of HTSC and the Mott's J, which is  derived from this study, may contribute to a better understanding of superconductivity in hole-doped cuprates.\\
%%%%%%%%%%%%%%%%%%%%%%%%%%%%%%%%%%%%%%%%%%%%%%%%%%%%%%%%%%%%%%%%%%%%%%%%%%%%%%%%%

\section{The Bibliography}
%Bibliographies are very important in Junior Lab papers.  Beyond the
%requisite citation of source material, they provide
%evidence of your investigations beyond the narrow scope of the
%labguide, something explicitly required of all Junior
%Lab students!  Good bibliograhies are doubly important in the real
%world where they are very (often the most) important sources
%of information for researchers entering the field.  Bibliographic
%entries may be made either in the `.tex' file itself or within a
%separate `.bib' file which gets attached during process of building a
%final PDF document.  This latter method is the preferred method and is
%then one used in this template by default.  An example of the
%alternative style, currently commented out,  is contained in the `.tex' source file.
%%%%%%%%%%%%%%%%%%%%%%%%%%%%%%%%%%%%%%%%%%%%%%%%%%%%%%%%%%%%%%%%%%%%%%%%%%%%%%%%%%
% Place all of the references you used to write this paper in a file
% with the same name as following the \bibliography command
%%%%%%%%%%%%%%%%%%%%%%%%%%%%%%%%%%%%%%%%%%%%%%%%%%%%%%%%%%%%%%%%%%%%%%%%%%%%%%%%%%
\bibliography{sample-paper}
\bibliographystyle{prsty}

%%%%%%%%%%%%%%%%%%%%%%%%%%%%%%%%%%%%%%%%%%%%%%%%%%%%%%%%%%%%%%%%%%%%%%%%%%%%%%%%%%
%%%%%%%%%%%%%%%%%%%%%%%%%%%%%%%%%%%%%%%%%%%%%%%%%%%%%%%%%%%%%%%%%%%%%%%%%%%%%%%%%%
%\begin{acknowledgments}I would like to thank all the members of the Physical Society of Japan for their constant inspiration.
%and Dr.Thomas Busch, Professor of Quantum Physics,
%\end{acknowledgments}
%%%%%%%%%%%%%%%%%%%%%%%%%%%%%%%%%%%%%%%%%%%%%%%%%%%%%%%%%%%%%%%%%%%%%%%%%%%%%%%%%%
%%%%%%%%%%%%%%%%%%%%%%%%%%%%%%%%%%%%%%%%%%%%%%%%%%%%%%%%%%%%%%%%%%%%%%%%%%%%%%%%%%
%\clearpage
\appendix

\end{document}